# Structural biology in the clouds: The WeNMR-EOSC Ecosystem

Rodrigo Vargas Honorato[1], Panagiotis I. Koukos[1], Brian Jiménez-García[1], Andrei Tsaregorodtsev[2], Marco Verlato[3], Andrea Giachetti[4], Antonio Rosato[4] and Alexandre M.J.J. Bonvin[1*]

[1]Bijvoet Centre for Biomolecular Research, Faculty of Science, Department of Chemistry, Utrecht University, Padualaan 8, 3584CH Utrecht, Netherlands

[2]Aix Marseille University, CNRS/IN2P3, CPPM, Marseille, France

[3]INFN, Sezione di Padova, Via Marzolo 8, 35131 Padova, Italy

[4]Department of Chemistry and Magnetic Resonance Center, University of Florence, and C.I.R.M.M.P., Via Luigi Sacconi 6, 50019 Sesto Fiorentino, Italy

**\*Correspondence:**
Corresponding Author
a.m.j.j.bonvin@uu.nl

**Abstract**

Structural biology aims at characterizing the structural and dynamic properties of biological macromolecules at atomic details. Gaining insight into three dimensional structures of biomolecules and their interactions is critical for understanding the vast majority of cellular processes, with direct applications in health and food sciences. Since 2010, the WeNMR project (www.wenmr.eu) has implemented numerous web-based services to facilitate the use of advanced computational tools by researchers in the field, using the high throughput computing infrastructure provided by EGI. These services have been further developed in subsequent initiatives under H2020 projects and are now operating as Thematic Services in the European Open Science Cloud (EOSC) portal (www.eosc-portal.eu), sending >12 millions of jobs and using around 4000 CPU-years per year. Here we review 10 years of successful e-infrastructure solutions serving a large worldwide community of over 23,000 users to date, providing them with user-friendly, web-based solutions that run complex workflows in structural biology. The current set of active WeNMR portals are described, together with the complex backend machinery that allows distributed computing resources to be harvested efficiently.

**Structural biology in the clouds**

# 1    INTRODUCTION

Proteins and nucleic acids are the main biological macromolecules responsible for function and maintenance of most of the machinery of life. The study of these macromolecules, their three-dimensional (3D) structure, dynamical behavior and interactions are crucial to gain a better understanding of relevant biological processes both related to biotechnological applications as well as health related such as the development of new drugs.

Since many years, in the context of WeNMR ([www.wenmr.eu](www.wenmr.eu)) now operating as a Thematic service provider in the European Open Science Cloud (EOSC), we have provided the community with relevant and specialized software that make up a valuable toolkit for structural biology researchers. One critical aspect aside from providing the community with valuable tools is to also provide the means of executing them in a user-friendly, efficient and cost-effective distributed manner, as done via the DIRAC Workload Manager (Tsaregorodtsev and Project, 2014).

Here we present an overview of the structural biology services provided via the WeNMR-EOSC ecosystem, what are the technological challenges and solutions that need to be implemented for an efficient use of distributed computing resources and how those are used by an active community worldwide.

# 2    SERVICES

The study of macromolecular structures is a multifaceted challenge in which researchers must apply different approaches and tools to gain a better understanding of their function, behavior and dynamics. To this end, within the WeNMR-EOSC project, we have been developing different tools, making them freely available to the community as webservices. The following paragraphs provide a short overview of currently active services under the EOSC portal.

## 2.1   AMPS-NMR

The AMPS-NMR (AMBER-based Portal Server for NMR structures) (Bertini et al., 2011) provides a user-friendly interface to run restrained molecular dynamics simulations (rMD) to refine experimental NMR structures, and to perform molecular dynamics simulations of biomacromolecular systems in general. Calculations can be run on CPUs or GPGPUs. The latter provide significantly faster performance (Andreetto et al., 2017) but are limited in availability. For NMR structures, the use of the AMPS-NMR portal results in a consistent improvement of features such as rotamer distributions, backbone normality and occurrence of steric clashes, without affecting agreement with the experimental data. The effect of rMD refinement is especially relevant for protein regions where the amount of experimental information is scarce. The rMD protocol has been implemented as a predefined multi-step procedure, so that the user does not need to know in detail how to tune the many parameters involved in a relatively complex MD calculation; the portal also handles automatically the most commonly used formats for experimental data. The AMPS-NMR webserver is available at http://py-enmr.cerm.unifi.it/access/index

## 2.2 DISVIS

The interaction between molecules can be experimentally characterized with methods such as cross-linking mass spectrometry (XL-MS) and FRET (Okamoto and Sako, 2017; Yu and Huang, 2017) which can provide information about which residues are participating in the interaction as well as the distance (or an upper limit to it) between the reacting groups of the residues. DISVIS is a tool that allows researchers to visualize the three-dimensional distance-restrained possible interaction space between the interacting proteins partners. It can be used to investigate if the restraints are mutually exclusive and identify possible false positives in the input data. DISVIS is freely available at https://wenmr.science.uu.nl/disvis.

## 2.3 FANTEN

FANTEN (Rinaldelli et al., 2015) aims at the determination of the anisotropy tensors associated with NMR paramagnetic pseudocontact shifts (pcs) and/or residual dipolar couplings (rdc's). FANTEN also permits the pcs-driven determination of the 3D structure of protein-protein adducts using a rigid body approach; the user can then submit these models directly to HADDOCK for flexible refinement. The FANTEN web server is available at http://abs.cerm.unifi.it:8080/.

## 2.4 HADDOCK

The pioneer of integrative docking, HADDOCK (*High Ambiguity Driven biomolecular DOCKing*) (Dominguez et al., 2003) allows users to make use of a variety of experimental and theoretical information to drive the docking of up to 20 macromolecules. The latest version (v2.4) supports the docking of proteins, small-molecules, nucleic acids, peptides, cyclic peptides, glycans and glycosylated proteins. To accommodate the complexity that naturally arises from >18 years of development of new features and docking modes without compromising usability and accessibility, HADDOCK has been made freely available (with registration) to the community since 2008 via an intuitive webserver (Vries et al., 2010; Zundert et al., 2016). Since exposing 400+ parameters could hinder user experience (and usability), users are separated in access tiers: Users belonging to the "Easy" tier can access and modify a limited number of parameters

**Structural biology in the clouds**whereas users belonging to the "Guru" tier (granted upon specific request from users in their registration page) have full access and control of the docking protocol. The HADDOCK webserver is available at https://wenmr.science.uu.nl/haddock2.4

### 2.5   MetalPDB

MetalPDB (Andreini et al., 2013a; Putignano et al., 2017) is a database collecting information on the metal-binding sites present in the 3D structures of biological macromolecules (Putignano et al., 2017). In particular, metal sites are represented in MetalPDB, as 3D templates describing the local environment around the metal ion(s) independently of the overall macromolecular fold. Among its features, MetalPDB provides detailed statistical analyses regarding metal usage in proteins of known 3D structure, encompassing protein families, and in catalysis; it also includes tools for the structural comparison of metal sites (Andreini et al., 2013b). MetalPDB is available at https://metalpdb.cerm.unifi.it/

### 2.6   PDB-tools web

The classic Protein Data Bank (PDB) file format is a flat text file that is still used by many structural biology software to represent the spatial coordinates of macromolecular structures, even though the official format of the worldwide PDB (wwPDB) (Berman et al., 2003) has now shifted to the Crystallographic Information Framework (mmCIF) dictionary (Bourne et al., 1997). The simple flat text PDB format, despite its apparent simplicity, follows strict formatting rules which can be easily overlooked during editing resulting in various errors. PDB-tools web (Jiménez-García et al., 2021) is a fully configurable, user-friendly web interface for the command-line application *pdb-tools* (Rodrigues et al., 2018). Using the portal users can, in a few clicks, build a complex pipeline which can then be saved (and uploaded) for future use and reproducibility. The webserver is available at https://wenmr.science.uu.nl/pdbtools/.

### 2.7   Powerfit

Cryo-electron microscopy (cryo-EM) is an experimental technique that can be used to obtain direct images of large macromolecular complexes (Fernandez-Leiro and Scheres, 2016). When the resolution of the resulting EM data is not sufficient to build models directly in the maps, it is common to fit existing 3D structures into those maps. To this end, Powerfit has been developed as a Python package for fast and sensitive rigid body fitting of molecular structures into EM density maps with a core-weighted local cross correlation scoring function (Zundert and Bonvin, 2015b). It is freely available as a webserver (Zundert et al., 2017) running on both local and GPU-accelerated distributed infrastructures at https://alcazar.science.uu.nl/services/POWERFIT.

## 2.8 proABC-2

Monoclonal antibodies have been consolidated as therapeutic tools due to their high affinity and specificity towards their target. proABC-2 (Ambrosetti et al., 2020), the updated version of proABC (Olimpieri et al., 2013), is a valuable tool that predicts which antibody residues can form intermolecular contacts with its antigen and further give insight into the chemical components of such interactions, differentiating between hydrophobic and hydrophilic interactions. The latest version replaced the previously used random forest paratope predictor by a convolutional neural network algorithm. Predictions obtained by proABC-2 can be used for example to guide molecular docking, leading to improvements of success rate and quality of the predicted models. proABC-2 is available at https://wenmr.science.uu.nl/proabc2/.

## 2.9 Prodigy

In addition to elucidating the structural aspects of a biomolecular complex and its interacting components, it is of crucial importance to be able to estimate the strength of the interaction, or, in other words, to be able to predict the binding affinity of the complex. PRODIGY, the *PROtein binDIng enerGY Prediction tool* (Vangone and Bonvin, 2015) allows to predict the binding affinity of protein-protein complexes from their three-dimensional structures. It does so by calculating features derived both from interfacial residue contacts as well as those from the non-interface residue (Kastritis et al., 2014). The same rationale has been expanded to protein-ligand complexes (PRODIGY-LIGAND) (Kurkcuoglu et al., 2018; Vangone et al., 2018) and to the classification interfaces in crystal structures as biological or crystallographic (PRODIGY-CRYSTAL) (Elez et al., 2018; Jiménez-García et al., 2019). PRODIGY services are available at https://wenmr.science.uu.nl/prodigy/

## 2.10 SpotON

Once an interaction has been characterized, it is valuable to know which parts of the interface contribute more to the binding, namely what are interaction "hot spots" - residues that upon mutation to alanine confer a difference in the binding free energy of the complex greater than 2.0 kcal/mol. The experimental identification of these regions is typically low throughput, laborious and expensive since it involves multiple rounds of point mutations and binding energy evaluations. SpotON was developed a computational approach to identify and classify interface residues as either hot spots, with positive contribution to the interface or null spots, with little to no significant contribution to the interface (Melo et al., 2016; Moreira et al., 2017). Its predictor combines different machine-learning algorithms, and it is available as a free and easy-to-use webserver https://alcazar.science.uu.nl/services/SPOTON.

**Structural biology in the clouds**

## 3 INFRASTRUCTURE

The WeNMR portals use either local or distributed computing resources harvesting is some case GPU resources to speed up the computations. Some portals, especially the ones making use of the EOSC-EGI high-throughput computational resources (see below) do require registration for use (Table 1).

**Table 1** – Execution infrastructure, registration policy and certificate handling for each WeNMR-EOSC service.

| Service | Infrastructure[1] | Registration required | Certificate Handling[2] | Reference |
|---|---|---|---|---|
| AMPS-NMR | Local/HTC/GPU | Yes | Robot | (Bertini et al., 2011) |
| DISVIS | Local/HTC/GPU | Yes | Robot | (Zundert and Bonvin, 2015a) |
| FANTEN | Local/HTC | No | No | (Rinaldelli et al., 2015) |
| HADDOCK | Local/HTC | Yes | Robot | (Vries et al., 2010) |
| MetalPDB | Local | No | No | (Putignano et al., 2017) |
| PDB-tools web | Local | No | No | (Jiménez-García et al., 2021) |
| Powerfit | Local/HTC/GPU | Yes | Robot | (Zundert et al., 2017) |
| proABC-2 | Local | No | No | (Ambrosetti et al., 2020) |
| Prodigy | Local | No | No | (Xue et al., 2016; Vangone et al., 2018; Jiménez-García et al., 2019) |
| SpotON | Local | Yes | No | (Moreira et al., 2017) |

1) Local: Runs on local resources; HTC: Makes use of the EOSC/EGI distributed high-throughput computing resources; GPU: Can make use of local of EOSC/EGI distributed GPU resources
2) Portals accessing EOSC/EGI HTC resources do require X509 certificates. Robot means that the portal is using a robot X509 certificate and users do not require personal certificates.

### 3.1 EOSC-EGI

The EGI Federation is an international e-Infrastructure providing advanced computing and data analytics for research and innovation. In the last decade it evolved from the high energy physics compute grid (WLCG) towards a multi-disciplinary, multi-technology infrastructure federating hundreds of resource centers worldwide and delivering large-scale data analysis capabilities (> 1 millions of CPU-cores and > 900 PB of disk and tape storage) to more than 70,000 researchers covering many scientific disciplines. During 2020 it delivered 13 billions of CPU hours of High Throughput Computing (HTC) and 18 millions of CPU hours of cloud computing.

EGI was one of the main actors and the coordinator of EOSC-Hub, a three-year European project that ended in March 2021, aimed at starting the design and implementation of the European Open Science Cloud. EGI operates several core elements of EOSC and contributes to EOSC service portfolio offering services like e.g., HTC, Cloud Compute, Workload Manager and Check-in which are relevant for WeNMR. Moreover, with the support of the recently started EGI-ACE project, under which WeNMR services are further operated, EGI will deliver the EOSC Compute Platform and will contribute to the EOSC Data Commons through a federation of cloud compute and storage facilities, Platform as a Service (PaaS) services and data spaces with analytics tools and federated access services.

The collaboration between EGI and WeNMR dates back to their early days, when in 2011 WeNMR became the first Virtual Research Community officially recognized by EGI. A formal Service Level Agreement (SLA) was established in 2016 and has since been periodically updated and recently extended until June 2023. The SLA between EGI and WeNMR is a document summarizing ten Operational Level Agreements (OLAs) between EGI and ten resource centers committed to provide HTC, cloud compute and storage capacity with availability and reliability above well-defined targets constantly monitored by the EGI operation team.  These ten resource centers are located in Czech Republic, Italy, Portugal, Spain, Taiwan, The Netherlands and Ukraine. They ensure the availability of 6050 HEPSPEC'06 normalized CPU-years per year of HTC computing time, 540 virtual CPU-cores of cloud compute capacity, and 59 TB of online storage capacity. Furthermore, a grid site hosted at the *Consorzio Interuniversitario Risonanze Magnetiche di Metalloproteine Paramagnetiche* (CIRMMP) provides several GPGPU cards that greatly enhanced the performance of the DISVIS, POWERFIT and AMPS-NMR applications.  Next to the SLA sites, additional resources are also available on an opportunistic use basis. During 2020 the HTC capacity has further increased by adding some Open Science Grid resources hosted in the US and a few WLCG sites hosted in France, Germany and Spain that decided to support WeNMR COVID-19 related jobs.

### 3.2   SSO

The WeNMR community has since the beginning recognized the importance of having a Single Sign-On (SSO) mechanism to access their web portals with unique login credentials. The first SSO was implemented in 2013 under the WeNMR European project, allowing users to subscribe to portals, validate personal grid certificates and manage job submissions all from one web page, collecting at the same time accounting records and user's geographic information. A few years later, with the support of the West-Life H2020 European project (2015-2018), the WeNMR SSO was gradually replaced with the West-Life SSO. In recent years the SSO mechanisms typically have relied on an Authentication and Authorization Infrastructure (AAI) based on services like the IdP-SP-Proxy and standard protocols like SAML, OpenID Connect and OAuth, and have been designed following the guidelines of the *Authentication and Authorization for Research and Collaboration* (AARC) initiative launched in May 2015. With the support of the EOSC-Hub project (2018-2021) the WeNMR services transitioned to the current AAI solution based on the EGI

Check-in service. Developed from the same AARC Blueprint Architecture[1], EGI Check-in solution is part of EOSC service catalogue, and has also been adopted by the Instruct-ERIC infrastructure. Nowadays all the WeNMR portals integrate EGI Check-in as SSO mechanism. Users can register to the EGI Check-in at https://aai.egi.eu and use those credentials to register to WeNMR services. All portals are compliant with the European General Data Protection Regulation (GDPR) with clear condition of use (https://wenmr.science.uu.nl/conditions) and privacy measures (https://wenmr.science.uu.nl/privacy).

### 3.3   DIRAC Workload Manager

The EGI Workload Manager is a service enabling users to access distributed computing resources of various types through a single user-friendly interface. It is one of the services offered by the EGI e-infrastructure project and it is also available via the EOSC Compute Platform. The service is built upon the software from the DIRAC Interware project (http://diracgrid.org) (Tsaregorodtsev and Project, 2014). The project is providing a development framework and a set of ready-to-use components to build distributed computing systems of arbitrary complexity. It provides a complete solution for communities needing access to computing and storage resources distributed geographically, integrated in different grid and cloud infrastructures or standalone computing clusters and supercomputers. The DIRAC Workload Manager serves multiple scientific communities – partners of the EGI e-infrastructure and the WeNMR Collaboration as one of its most active users. User tasks prepared by the WeNMR application portals are submitted to the Workload Manager service. The service performs reservation of computing resources by means of so-called pilot jobs which are submitted to various computing centers with appropriate access protocols. Once deployed on worker nodes, pilot jobs verify the execution environment and then request user payloads from the central DIRAC Task Queue. Altogether, the pilot jobs and the central Task Queue form a dynamic virtual batch system that overcomes the heterogeneity of the underlying computing infrastructures. This workload scheduling architecture allows to easily add new resources transparently for the users. It also makes it easy to apply usage policies by defining fine grained priorities to certain activities. For example, during the COVID pandemic it allowed to quickly make available resources of the sites willing to contribute to the COVID-related studies and ensure high priority of these jobs compared to other regular WeNMR activities.

### 4   COMMUNITY AND USAGE

The WeNMR services combined have had over 150.000+ users (Table 2). User satisfaction is monitored and the services have various support mechanisms summarized at

---

[1] https://aarc-project.eu/wp-content/uploads/2019/05/AARC2-DJRA1.4_v2-FINAL.pdf

**Structural biology in the clouds**

[https://www.wenmr.eu/support](https://www.wenmr.eu/support) allowing users to fill in requests and ask specific questions. This large userbase is provided with tutorials to ensure the tools are being used to their best (Table 3). The user's feedback and issues are one of the driving forces towards the improvement of the services and implementation of new features.

**Table 2** – Usage statistics and satisfactions of the WeNMR services

| SERVICE | SATISFACTION RATING [SCALE 1-5] (#OF RESPONDENTS) | TOTAL NUMBER OF USERS TO DATE | TOTAL NUMBER OF SERVED REQUESTS TO DATE | WEB PORTAL RELEASE YEAR |
|---|---|---|---|---|
| AMPS-NMR | - | 657 | 27 144 | 2011 |
| DISVIS | 4.76 (75) | 3 292 | 3 823 | 2016 |
| HADDOCK | 4.9 (5380) | 22 189 | 342 755 | 2008 |
| METALPDB | - | 158 930* | 225 425** | 2013 |
| PDB-TOOLS WEB | 4.91 (74) | - | 5 012 | 2020 |
| POWERFIT | 4.75 (16) | 2 622 | 701 | 2016 |
| PROABC-2 | 4.84 (73) | - | 7 682 | 2019 |
| PRODIGY | 4.78 (1 991) | - | 136 420 | 2016 |
| SPOTON | 4.7 (82) | 3 013 | 6 700 | 2017 |

\* UNIQUE IPS
\*\* PAGES SERVED

**Table 3** – Tutorials and user support material for the WeNMR-EOSC services

| Service | Tutorials/Manuals | Support |
|---|---|---|
| AMPS-NMR | https://www.wenmr.eu/tutorials/#rmd-amber | Via e-mail |
| DISVIS | https://www.bonvinlab.org/education/Others/disvis-webserver/ | https://ask.bioexcel.eu |
| FANTEN | http://abs.cerm.unifi.it:8080/manual/FANTEN_manual.pdf | Via e-mail |
| HADDOCK | https://www.bonvinlab.org/education/HADDOCK24/ | https://ask.bioexcel.eu |
| MetalPDB | https://metalpdb.cerm.unifi.it/glossary | Via e-mail |
| PDB-tools web | https://wenmr.science.uu.nl/pdbtools/manual | https://ask.bioexcel.eu |
| Powerfit | https://www.bonvinlab.org/education/Others/powerfit-webserver | https://ask.bioexcel.eu |
| proABC-2 | https://wenmr.science.uu.nl/proabc2/manual | Via e-mail |
| Prodigy | https://wenmr.science.uu.nl/prodigy/manual | https://ask.bioexcel.eu |
| SpotON | https://alcazar.science.uu.nl/cgi/services/SPOTON/spoton/help | https://ask.bioexcel.eu |

**Structural biology in the clouds**

The services discussed here have been making use of EGI HTC resources since 2009. Figure 1, plots the normalized CPU usage per month, clearly showing the increasing use of the EGI HTC resources over time, culminating in 2020 to the equivalent of ~5300 CPU years' worth of computing.

During the COVID pandemic, one of the most used WeNMR service, HADDOCK, has seen a significant increase if both the number of submissions and single users, with about 1/3 of all submissions since April 2021 being COVID-related (Figure 2).

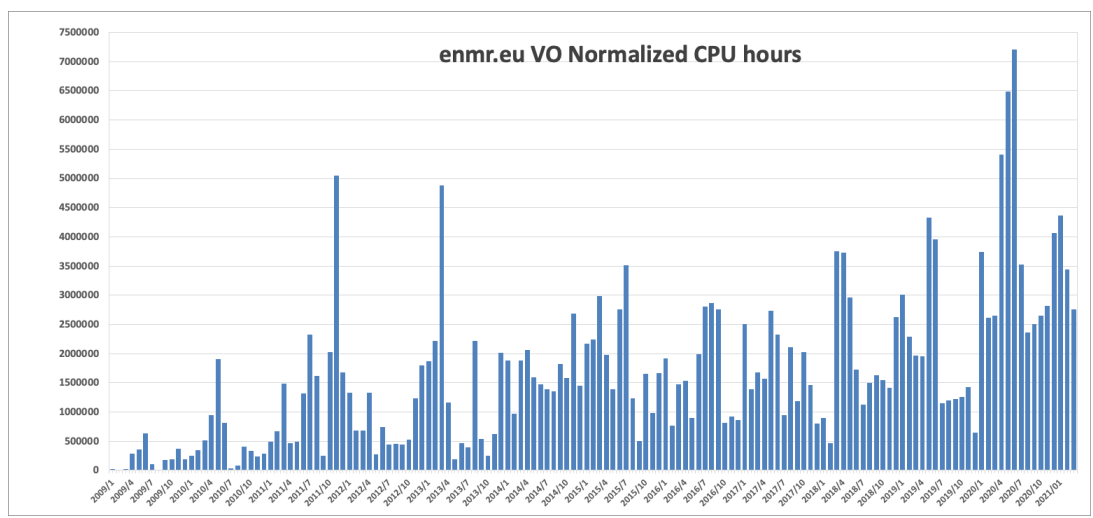

**Figure 1**: Normalized CPU use per month (in CPU hours) of the WeNMR services (source EGI accounting portal).

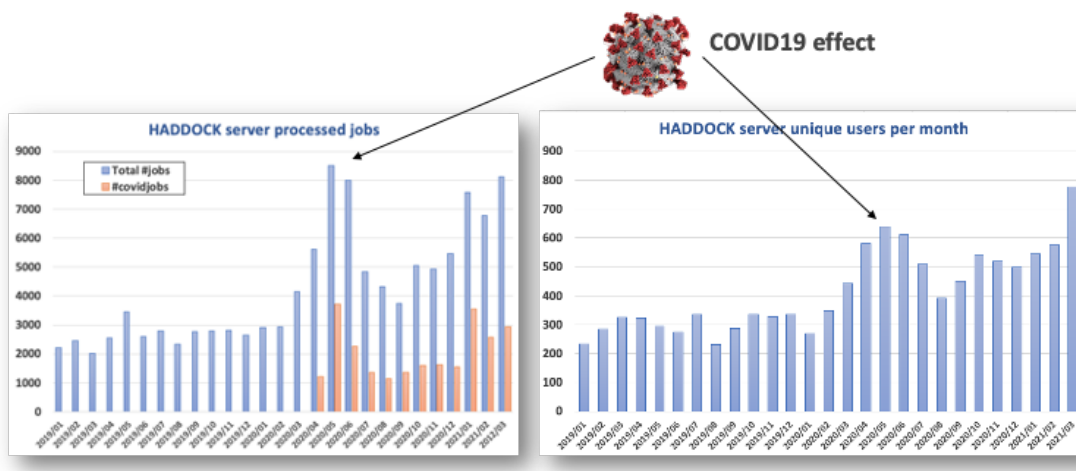

**Figure 2**: Number of submissions (left) and single users (right) per month of the HADDOCK WeNMR portal since 2019. Since April 2020 users can flag their submission as COVID-related (orange bars). Those represent about 1/3 of the total number of submissions.

## 5      Conclusions

Over more than a decade the WeNMR project has facilitated the use of advanced computational tools. Taking advantage of the EGI high throughput compute infrastructure, it has developed as a Thematic Services in the European Open Science Cloud. Its large worldwide user community of over 23,000 users distributed over 125 countries, submitted in 2020 >12 million jobs that accounted for ~4,000 CPU-years. In tight collaboration with EGI, WeNMR has constantly improved its services, for example piloting the use of distributed GPU resources and facilitating user access through the implementation of a single sign one mechanism. Its valuable services are offered via user-friendly web-based solutions that allows researchers to execute complex workflows at the click of their mouse without having to deal with the complexity of managing and distributing computations. The impact is evident from both usage of the services in research and education and from the number of citations of the various tools (>800 since 2020). The active WeNMR user community, in turn, acts a catalyst for further development and improvement of the services.

## 6      Conflict of Interest

None.

## 7      Author Contributions

RVH and AMMJB conceived the structure of the manuscript and all authors contributed to the writing.

## 8      Funding

This work is co-funded by the Horizon 2020 projects EOSC-hub (grant number 777536), EGI-ACE (grant number 101017567) and BioExcel (grant numbers 823830 and 675728) and by a computing grant from NWO-ENW (project number 2019.053).

## 9      Acknowledgments

The EGI team over the years is acknowledged for their continuous support in the operation of our services and facilitating the access to the EOSC/EGI HTC resources.

# Structural biology in the clouds